# The Effect of Polar Fluctuation and Lattice Mismatch on Carrier Mobility at Oxide Interfaces

Z. Huang[1,*], K. Han[1,2,*], S. W. Zeng[1], M. Motapothula[1], W. M. Lü[1], C. J. Li[1], W. X. Zhou[1,2], J. M. D. Coey[1,3], T. Venkatesan[1,2,4,§], and Ariando[1,2,§]

[1]*NUSNNI-NanoCore, National University of Singapore, Singapore 117411, Singapore.*

[2]*Department of Physics, National University of Singapore, Singapore 117542, Singapore.*

[3]*School of Physics and CRANN, Trinity College, Dublin 2, Ireland.*

[4]*Department of Electrical and Computer Engineering, National University of Singapore, Singapore 117576, Singapore.*

*These authors contributed equally to this work.*

[§]*Correspondence and requests for materials should be addressed to Ariando (email: ariando@nus.edu.sg) or TV (email: venky@nus.edu.sg)*

**Since the discovery of two-dimensional electron gas (2DEG) at the oxide interface of $LaAlO_3/SrTiO_3$, improving carrier mobility has become an important issue for device applications. In this paper, by using an alternate polar perovskite insulator $(La_{0.3}Sr_{0.7})(Al_{0.65}Ta_{0.35})O_3$ (LSAT) for reducing lattice mismatch from 3.0% to 1.0%, the low-temperature carrier mobility has been increased 30 fold to 35,000 $cm^2V^{-1}s^{-1}$. Moreover, two critical thicknesses for the $LSAT/SrTiO_3$ (001) interface are found: one at 5 unit cell for appearance of the 2DEG, the other at 12 unit cell for a peak in the carrier mobility. By contrast, the conducting (110) and (111) LSAT/STO interfaces only show a single critical thickness of 8 unit cells. This can be explained in terms of polar fluctuation arising from LSAT chemical composition. In addition to lattice mismatch and crystal symmetry at the interface, polar fluctuation arising from composition has been identified as an important variable to be tailored at the oxide interfaces to optimise the 2DEG transport.**

Strongly-correlated electrons in oxide heterostructures can exhibit various remarkable properties due to the mismatch of lattice, polarization, composition and orbital character at the interface [1]. The most famous example is the high-mobility two-dimensional electron gas (2DEG) at the interface between the two insulators SrTiO$_3$ (STO) and LaAlO$_3$ (LAO) [2], which can exhibit two-dimensional superconductivity [3], magnetic interactions [4], and electronic phase separation [5-8]. The appearance of such a 2DEG is often ascribed to the polar discontinuity arising at the interface between the polar LAO overlayer and a nonpolar STO (001) substrate [2,9-11]. An internal electric potential $V$ can be built up in the alternating stack of polar AlO$_2^-$/LaO$^+$ layers on the TiO$_2^0$-terminated nonpolar STO. When $V$ exceeds the STO bandgap $E_{g,STO}$ (= 3.2 eV), which is much less than that of LAO (= 5.6 eV), electrons can be transferred from the LAO valance band to the STO conduction band. This electronic reconstruction creates the 2DEG at the STO side of the LAO/STO interface. The minimal thickness of the polar layer $t_C$ that is required for electronic reconstruction is $t_C = \varepsilon_0\varepsilon_P\Delta E/eP$, where $\varepsilon_P$ is the dielectric constant of the polar material, $\Delta E$ is the energy gap separating the valance band of the polar layer and the conduction band of the nonpolar material, and $P$ is the electric polarization of polar layers [10]. Taking $\varepsilon_P$ = 24, $\Delta E$ as STO bandgap of 3.2 eV, and $P$ = 0.526 C m$^{-2}$ for the LAO/STO (001) interface, $t_C$ is calculated to be 4 unit cells (uc) in perfect agreement with the experimental value [12]. Besides the polar discontinuity, recent experimental results also suggest that other factors, like oxygen vacancy [13,14] or chemical stoichiometry [15,16], can influence the SrTiO$_3$-based 2DEG, showing the complexity on engineering the oxide interfaces.

A focus of current research is to improve the carrier mobility, in order to make these conducting oxide interfaces more suitable for probing quantum transport [17-19] and for eventual electronic device application [20]. The electron mobility for the conventionally prepared LAO/STO interface is usually around 1,000 cm$^2$V$^{-1}$s$^{-1}$ at low temperatures [5,21,22], and it can be increased to 6,000 cm$^2$V$^{-1}$s$^{-1}$ by optimizing the growth condition [17]. An SrTiO$_3$/SrCuO$_2$ cap layer on the LAO/STO heterostructure can improve the carrier mobility up to 50,000 cm$^2$V$^{-1}$s$^{-1}$, provided the sample is prepared under high vacuum (~ 10$^{-6}$ mbar) [21]. Replacing the LAO overlayer with spinel γ-Al$_2$O$_3$ has been found to greatly increase the mobility to 140,000 cm$^2$V$^{-1}$s$^{-1}$, but again the high-mobility state collapses after thermal treatment in oxygen [23]. Therefore, these high-mobility 2DEG (> 10,000 cm$^2$V$^{-1}$s$^{-1}$) processes seem incompatible with other (hole-doped) functional perovskite oxides like superconducting cuprates or ferromagnetic manganites, for which an oxygen-rich environment is required for film growth. Hence, obtaining robust high carrier mobility at the conducting oxide interfaces, which can survive subsequent oxygen-rich processing condition remains a challenge.

Here we show that this can be accomplished by replacing LAO with LSAT – (La$_{0.3}$Sr$_{0.7}$)(Al$_{0.65}$Ta$_{0.35}$)O$_3$ or (LaAlO$_3$)$_{0.3}$(SrAl$_{0.5}$Ta$_{0.5}$O$_3$)$_{0.7}$. LSAT is a well-known cubic perovskite insulator with lattice constant $a_{LSAT}$ = 3.868 Å, dielectric constant ε$_{P,LSAT}$ = 22, and bandgap $E_{g,LSAT}$ = 4.9 eV [24]. When grown on STO, the lattice mismatch for LSAT/STO is only 1.0%, which is only one third of the value of LAO/STO (3.0%). Furthermore, STO and LSAT both undergo a similar cubic-to-tetragonal transition below 100 K [25,26], whereby maintaining the structural coherency. Figure 1(a) shows the typical *in-situ* Reflection High-Energy Electron Diffraction (RHEED) during the fabrication of the LSAT/STO (001) interface by pulsed laser deposition method [27]. Both the periodic RHEED oscillation and streaky RHEED pattern clearly demonstrate the layer-by-layer growth for LSAT/STO (001) interface. After growth, samples were *ex-situ* annealed for 1 hour at 600 $^o$C in 1 bar oxygen to remove the oxygen vacancy [14]. Taken together with the X-Ray reflectivity data (Fig. S1 in *Supplementary Materials* [28]), we can identify one RHEED oscillation with the growth of one perovskite unit cell. The *ex-situ* Atomic Force Microscopy (AFM) image and height profile in Fig. 1(b) also exhibit one-unit-cell-high steps on the LSAT/STO (001) surface. Rutherford Backscattering Spectrometry (RBS) (Fig. S2 in *Supplementary Materials* [29]) confirms that the chemical composition for the LSAT film is very close to the target composition $(La_{0.3}Sr_{0.7})(Al_{0.65}Ta_{0.35})O_3$, leading to an average polar charge density of $\pm 0.3e$ per $AO/BO_2$ layer along [001] axis, as illustrated in Fig. 1(c). Given the nonpolar nature of $SrTiO_3$ (001), a polar-discontinuity-induced 2DEG is expected at the LSAT/STO (001) interface.

The thickness-dependent transport data at 2 K for the LSAT/STO (001), (110), and (111) interfaces are summarized in Figs. 2(a)-2(c), respectively. In Fig. 2(a), the (001) interface becomes conducting when covered by a LSAT layer with thickness $t \geq 5$ uc. Furthermore, the low-temperature sheet conductance of LSAT/STO increasers with *t*, reaching its highest value at $t \approx 12$ uc. This conductance improvement is not caused by any increases of carrier density, but it is brought about by a great enhancement of carrier mobility $\mu_S$, which reaches its peak of 35,000 $cm^2V^{-1}s^{-1}$ at $t = 12$ uc. And this high carrier mobility is about 30 times larger than that of LAO/STO interfaces prepared under similar conditions [5,21,22]. Also, we note that clear Shubnikov-de Haas conductance oscillations can be observed at 2 K for (001) interfaces with high carrier mobility (Fig. S3 in *Supplementary Materials* [30]). In addition, the transport data clearly show that there are two critical thicknesses for the LSAT/STO (001) interface: one is at 5 uc where the 2DEG is established and the other is around 12 uc where the mobility is greatest.

On the other hand, this high mobility 2DEG is also observed at the annealed (110)- and (111)-orientated LSAT/STO interfaces, which is similar to the LAO/STO interface with different orientations [22,31]. But unlike LSAT/STO (001) interface, both the (110) and (111) interfaces show only a single critical thickness, at 8 uc, as shown in Figs. 2(b) and 2(c). At 2 K, the carrier density for all three annealed LSAT/STO interfaces is around $1-2\times10^{13}$ cm$^{-2}$, and the carrier mobility for (110) and (111) interfaces is around 6,000 cm$^2$V$^{-1}$s$^{-1}$. When compared to LAO/STO interfaces, the (110) and (111) LSAT/STO interfaces show much more robust metallicity. For example, our data show that the high-mobility 2DEG can be maintained in the LSAT/STO (110) and (111) interfaces with a 50-uc-thick LSAT layer, while the LAO/STO (110) and (111) interfaces show low-temperature insulating behavior when LAO thickness is beyond just 10 uc [22].

Therefore, two major differences between LSAT/STO and LAO/STO interfaces can be found in Fig. 2. *One* is the much higher carrier mobility and more robust metallicity at the LSAT/STO interface, and *the other one* is the observation of two critical thicknesses at (001) LSAT/STO interface, but not at the (110) and (111) LSAT/STO or all LAO/STO interfaces.

This high carrier mobility and robust metallicity of LSAT/STO can be ascribed to the small structural mismatch between the oxides. Such structural mismatch includes the nominal lattice mismatch calculated from lattice parameters, and the crystal symmetry mismatch related to octahedral rotation/tilting. For the conventional LAO/STO interface, the lattice mismatch is 3.0% at room temperature, three times of that of the LSAT/STO interface. This large lattice mismatch can induce distortion of the $B$O$_6$ octahedra creating structural defects to lower the carrier mobility near the LAO/STO interface, especially when LAO layer is thick. Also, the LAO/STO interface suffers from a rhombohedral/cubic symmetry mismatch, whereas both the LSAT and STO are cubic. So, there is no crystal symmetry mismatch and the octahedral tilting expected in the few STO layers near the LAO/STO interface should not arise for LSAT. In the case of CaTiO$_3$ (despite a lower lattice mismatch with LAO), the orthorhombic, octahedral tilting at the LaAlO$_3$/CaTiO$_3$ interface reduces the Ti-O-Ti bond angle and results in carrier localization [32]. Furthermore, the structural compatibility between STO and LSAT is maintained at low temperatures, because both oxides undergo a cubic-to-tetragonal structural transition below about 100 K [25,26]. Hence, by replacing LSAT with LAO, the structural mismatch is greatly reduced at the interface, resulting in the observed high carrier mobility at the LSAT/STO interfaces.

Following this idea, we can anticipate that the transport difference between LSAT/STO and LAO/STO, arising from the structural mismatch at the interface, will be even larger when the LAO and LSAT films are thicker. In Fig. 3, the LSAT/STO and LAO/STO interfaces with a thick polar overlayer (~ 100 uc) are compared. Figure 3(a) shows that the 100 uc LSAT/STO interface can still preserve the metallic behavior with $\mu_S$ as high as 6,500 $cm^2V^{-1}s^{-1}$ at 2 K, while the LAO/STO interface becomes insulating below 50 K. This suppression of conducting 2DEG in thick LAO/STO samples is also reported elsewhere [22,33]. In Fig. 3(b) of XRD $\theta$-$2\theta$ scans, well-defined thickness fringes can be observed for LSAT/STO interface, but no such fringes are seen for the LAO/STO interface. Moreover, by comparing Reciprocal Space Mapping (RSM) around the (103) reflection in Figs. 3(c) and 3(d), the structural reflection of LAO/STO is much worse than that of LSAT/STO. The LAO (103) reflection is diffusive and is divided into two parts, one of which, indicated by red arrow, shows structural relaxation tendency of the LAO to its bulk value. By contrast, the 100 uc LSAT/STO interface can still maintain the coherent growth with a sharp LSAT (103) reflection, showing less structural mismatch and hence higher carrier mobility. Furthermore, the lattice mismatch at the interface can also explain the suppression of carrier mobility in the thicker samples. As shown in the inset of Fig. 3(d), LAO/STO samples (5-25 uc) [33] show a faster decrease of mobility with thickness than LSAT/STO samples (12-25 uc), but both of them exhibit a similar tendency of mobility suppression with thickness. This drop of mobility with thickness should hence be ascribed to the lattice mismatch strain effect.

Now, we want to explain why the conducting LSAT/STO (001) interface exhibits two critical thicknesses, whereas for all LAO/STO, LSAT/STO (110) and (111) interfaces, there is only one. Based on the chemical composition, the LSAT lattice can be considered as consisting of 30% LAO and 70% $SrAl_{0.5}Ta_{0.5}O_3$ (SATO) sublattices as shown in Fig. 4(a). Note that these sublattices will have many possible arrangements and they are expected to be randomly arranged through out the thin film while maintaining the correct stoichiometry. However, if the LSAT is grown on a $TiO_2$-terminated STO (001), the electric polarization of the 30% LAO sublattice, $P_0$, will point from the surface to the interface, and can be calculated to be 0.523 $C/m^2$. On the other hand, for the 70% SATO sublattice, the electric polarization can be either parallel or anti-parallel to $P_0$ depending on the position of Al and Ta ($\pm P_0$ with an equal probability), leading to an average polarization of $0.3P_0$ for LSAT. Note that many possible combinations of the sublattices are possible while maintaining the correct stoichiometry of the LSAT compound, and this leads to different sublattice coloumns with a different polarization within the LSAT layer. Figure 4(b) shows one possible combination for a 6 uc LSAT layer on $TiO_2$-terminated STO (001), for which the polarization is $P_0$ in *Column A*, $0.3P_0$ in *Column B*, and $-0.4P_0$ in *Column C*, and the ratio of possibility among *Column A*, *B*, and *C* is 1:2:1. It can be seen that the LSAT (001) exhibits *polar fluctuation*- various values of polarization *P* for different sublattices. Based on the polar discontinuity model [9], electrons (holes) must be created to compensate $P_0$ ($-P_0$). Hence, a mobile 2DEG will be created in STO under *Column A* and *B*; while localized holes will exist under *Column C*, which can scatter the mobile electrons thereby lowering carrier mobility.

A statistical model is built to evaluate the polar fluctuation of LSAT/STO (001) interface, by assuming that one LSAT (001) monolayer could have 65% chance for producing an electric dipole moment as $P_0V$, or 35% chance for $-P_0V$, where V is the volume for one LSAT (001) monolayer. By applying the binomial distribution [34], the mean value of polarization $P_\mu$ is $0.3P_0$, and the standard deviation of polarization $\sigma_P$ is $(0.91/t)^{0.5}P_0$, where $t$ is the LSAT thickness (uc). Hence, the LSAT polarization $P$ can be characterized by $P_\mu \pm \sigma_P$ with a polar fluctuation $\sigma_P$, which varies with LSAT film thickness. On other words, the actual value of the polarization will lie between $P_\mu-\sigma_P$ and $P_\mu+\sigma_P$. If we define the critical polarization $P_C$ as the minimal polarization required for the polar layer with a given thickness $t$ to create 2DEG on STO, $P_C$ can be calculated by $P_C = \varepsilon_0\varepsilon_P E_{g,STO}/et$. As shown in Fig. 4(c), for LSAT (001) with polarization $P = P_\mu \pm \sigma_P$, there is an intersection between $P_C$ and $P_\mu+\sigma_P$ between 4-5 uc. It indicates the LSAT polar layer can stabilize a 2DEG on STO (001) when its thickness is above 5 uc, consistent with our observation on the first critical thickness ($t_1$) at 5 uc for the appearance of the 2DEG. Moreover, as the thickness increases beyond 5 uc, $P$ has reduced negative polarization component leading to an increasing mobility, which peaks at 10-11 uc where the negative polarization goes to zero. Beyond 10-11 uc, $P$ is always positive and the mobility cannot increase further. This is consistent with the observed second critical thickness where the mobility peaks at 12 uc. Beyond 12 uc the mobility decreases due to lattice strain as shown in the inset of Fig. 3 (d), albeit at a slower rate compared to LAO/STO. By contrast, for (110) and (111) orientations, the direction of SATO polarization cannot be changed by switching Al and Ta position, always pointing from $O_2$ to Sr(Al,Ta)O layer along (110) orientation, and from $SrO_3$ to (Al,Ta) layer along (111) orientation. Hence, there is no positively-charged scattering center and only one critical thickness for 2DEG is observed in the (110) and (111) interface.

In summary, when the lattice mismatch of conducting oxide interface on STO is reduced from 3.0% to 1.0% and the symmetry mismatch is minimized by replacing the polar layer LAO with cubic LSAT, the carrier mobility of 2DEG can be greatly improved (~35,000 cm$^2$V$^{-1}$s$^{-1}$ at 2 K), almost 30 times larger than the conventional LAO/STO prepared under the same conditions (~ 1,000 cm$^2$V$^{-1}$s$^{-1}$). Further this mobility is robust under different oxygen processing conditions. Moreover, the observation of two critical thicknesses for the LSAT/STO (001) interface but not the (110) and (111) interfaces (one for the appearance of 2DEG at 5 uc and the other for optimum carrier mobility at 12 uc) can be ascribed to the polar fluctuation in LSAT (001). Further improvement in the carrier mobility of the 2DEG induced in STO by polar discontinuity of the interface is likely, if the structural mismatch and polar fluctuation can be avoided.

## Acknowledgments

We thank H. Hilgenkamp for the discussion. This work is supported by the Singapore National Research Foundation (NRF) under the Competitive Research Programs (CRP Award No. NRF-CRP 8-2011-06 and CRP Award No. NRF-CRP10-2012-02) and the NUS FRC (AcRF Tier 1 Grant No. R-144-000-346-112).

Fig. 1. (Color online) (a) *In-situ* RHEED oscillations of 50 uc LSAT/STO (001). The intensity is manually increased at 300 seconds. Insets are the streaky RHEED pattern after growth and structural scheme of perovskite LSAT. (b) Surface profile of 20 uc LSAT/STO (001). The step height is around 3.85 Å. Inset is the AFM image, where the surface profile was taken along line A-B. (c) Scheme of polar discontinuity at the LSAT/STO (001) interface.

Fig. 2. (Color online) Sheet conductance, $G_{Sheet}$, carrier density, $n_S$, and carrier mobility, $\mu_S$, at 2 K are shown as a function LSAT thickness for the LSAT/STO (001) interface in (a), (110) in (b), and (111) in (c).

Fig. 3. (Color online) (a) $R_S(T)$ curves for 100 uc LAO/STO (001) and LSAT/STO (001) samples. (b) XRD $\theta$-$2\theta$ scans around (002) for LAO/STO and LSAT/STO. (c) RSM around (103) for 100 uc LAO/STO (001) sample. (d) RSM around (103) for 100 uc LSAT/STO (001) sample. The red line indicates the coherent growth with the same in-plane lattice constants for the film and the substrate. The green line indicates the LAO bulk (fully relaxed). The red arrow in (c) indicates the structural relaxation in LAO film. The inset in (d) shows the carrier mobility as a function of film thickness for (001) LAO/STO (Blue) and LSAT/STO (Red). The data for LAO/STO is from Ref [33].

Fig. 4. (Color online) (a) The LSAT layer contains 30% LAO sublattice ($P_0$) and 70% SATO sublattice ($\pm P_0$) on TiO$_2$-terminated STO (001). (b) Different polarizations for different columns, which are formed by randomly mixing 30% LAO and 70% SATO sublattices in the 6 uc LSAT/STO (001), leading

to 2DEG in STO beneath for *Column A* and *B*, and localized positively-charged holes for *Column C*. (c) Based on the binomial distribution, the LSAT polarization ***P***, which is characterized by ***P***$_\mu$ ± *σ*$_P$, is shown as a function of LSAT thickness. When the LSAT layer is 5 uc ($t_1$), LSAT polarization ***P*** with a positive fluctuation ***P***$_\mu$ + *σ*$_P$ is beyond the ***P***$_C$, starting to form 2DEG at the interface. When the LSAT layer is above 11 uc ($t_2$), LSAT polarization ***P***$_\mu$ − *σ*$_P$ is above zero, indicating the absence of the localized holes that scatter the 2DEG carriers.

**Figure 1:**

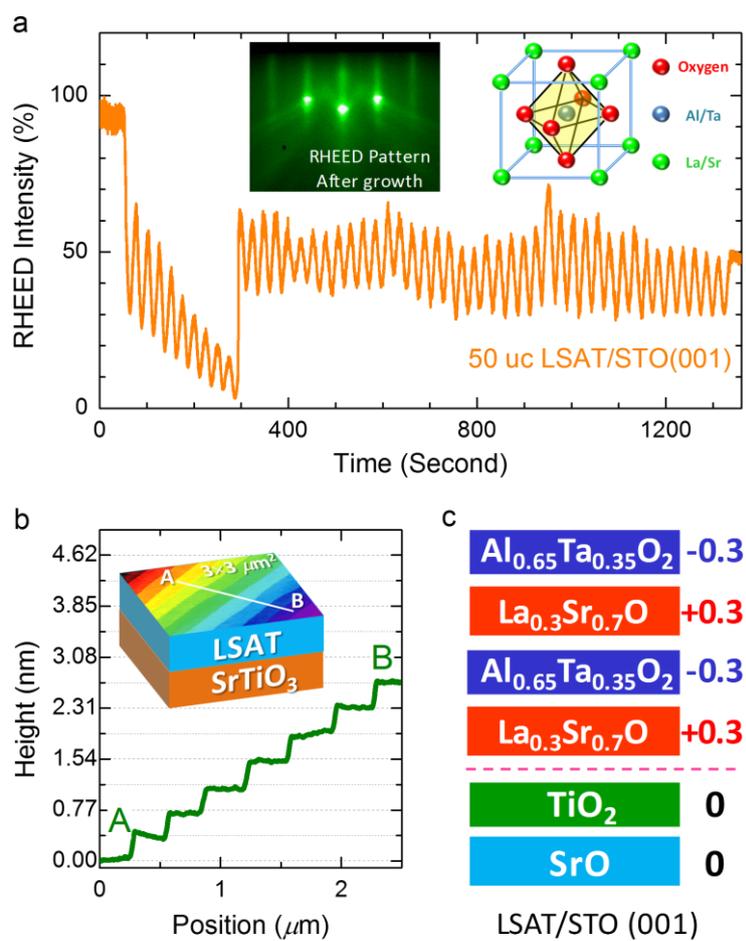

**Figure 2:**

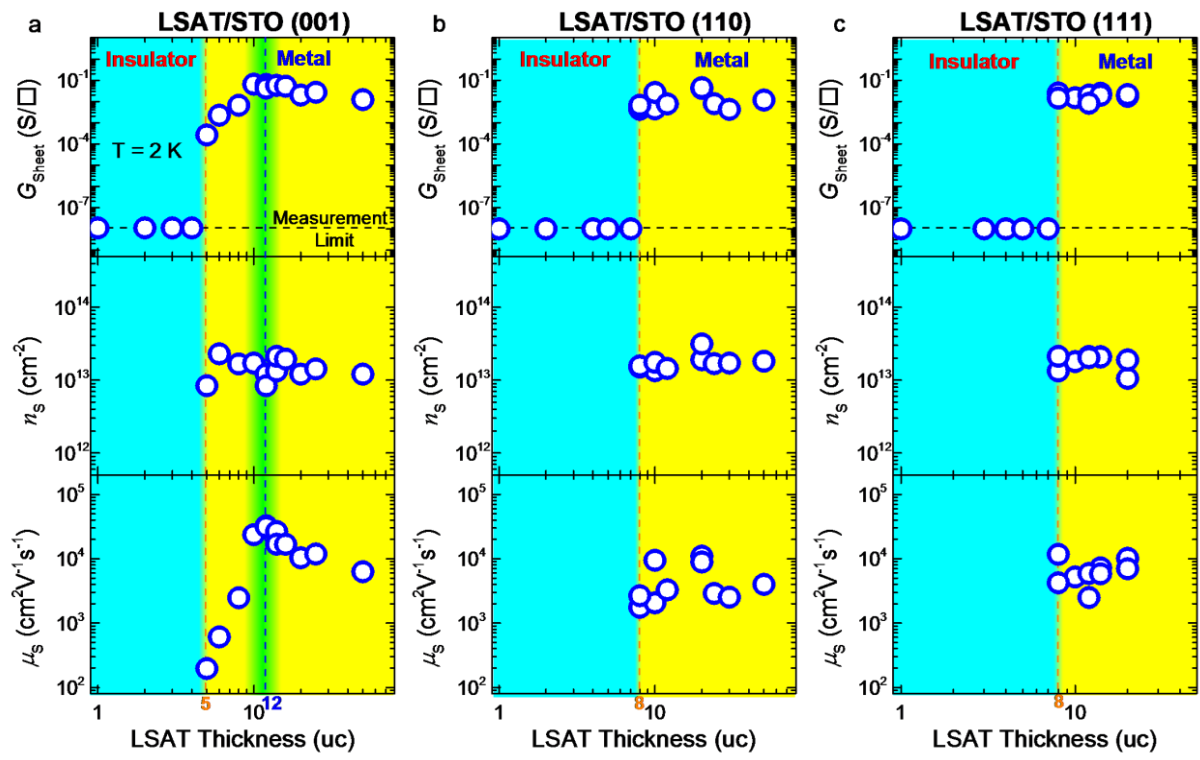

**Figure 3:**

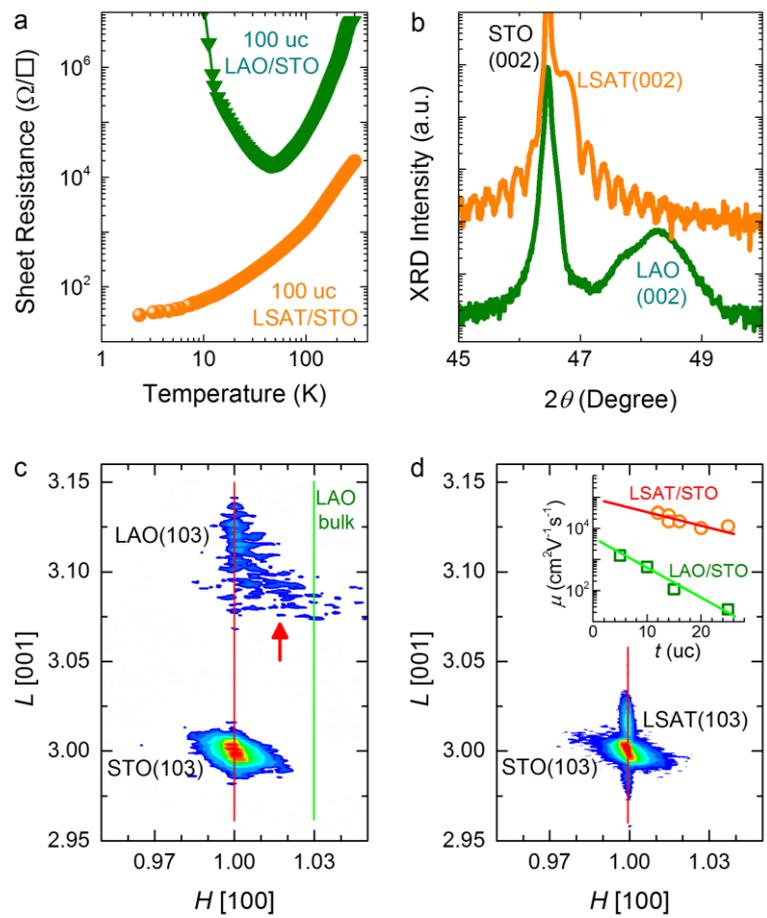

**Figure 4:**

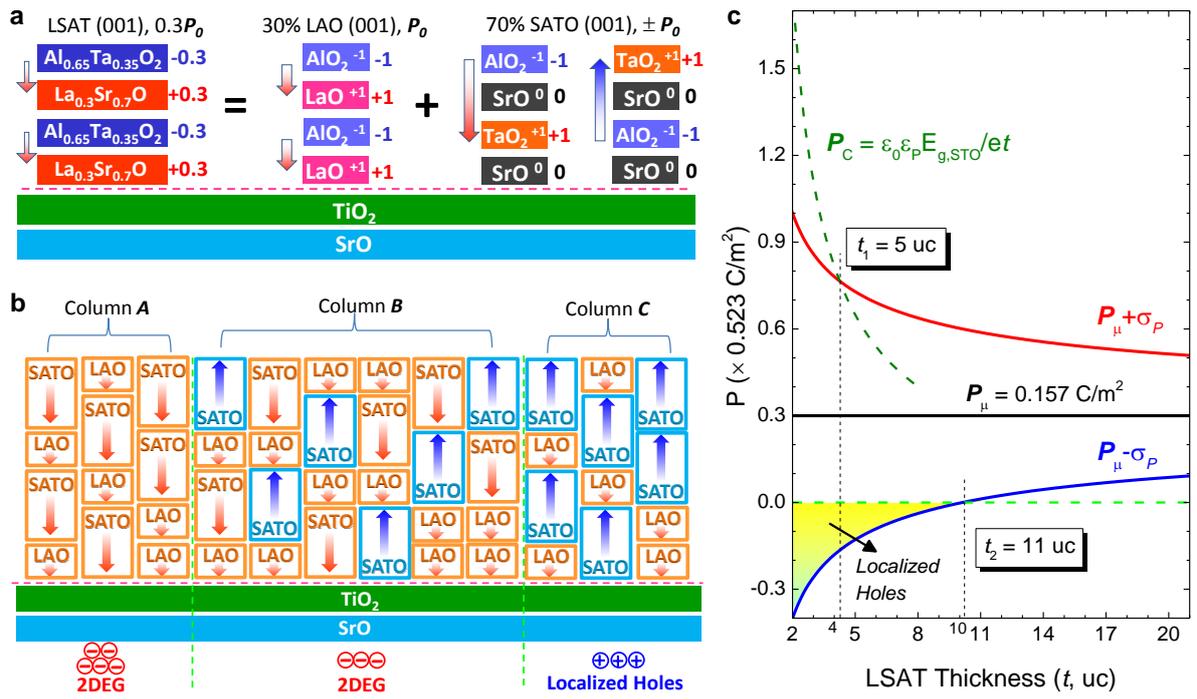